\documentclass[aps,amsfonts,reprint,tightenlines,amssymb,superscriptaddress,twocolumn,pra]{revtex4-1}

\usepackage[T1]{fontenc}
\usepackage{amsmath,amssymb,amstext,amsfonts,amsxtra}
\usepackage{graphicx}

\usepackage{times}
\usepackage{braket}
\usepackage{color}

\def\idty{{\leavevmode\rm 1\mkern -5.4mu I}} 
\newcommand{\diag}{\textnormal{diag}}
\newcommand{\BS}{\textnormal{BS}}
\newcommand{\QR}{\textnormal{QR}}

\begin{document}

\title{Repeaters for Continuous Variable Quantum Communication}

\author{Fabian Furrer}\email{furrer@eve.phys.s.u-tokyo.ac.jp}
\affiliation{NTT Basic Research Laboratories, NTT Corporation,
3-1 Morinosato-Wakamiya, Atsugi, Kanagawa, 243-0198, Japan.}

\author{William J. Munro} \email[]{bill.munro@me.com}
\affiliation{NTT Basic Research Laboratories, NTT Corporation,
3-1 Morinosato-Wakamiya, Atsugi, Kanagawa, 243-0198, Japan.}

\begin{abstract}
Optical telecommunication is at the heart of today's internet and is currently enabled by the transmission of intense optical signals between remote locations. As we look to the future of telecommunication, quantum mechanics promise new ways to be able to transmit and process that information. 
Demonstrations of  quantum key distribution and quantum teleportation using multi-photon states have been performed, but only over ranges limited to one hundred kilometers. To go beyond this, we need repeaters that are compatible with these quantum multi-photon continuous variables pulses. Here we present a design for a continuous variable quantum repeaters that can distribute entangled and pure two-mode squeezed states over arbitrarily long distances with a success probability that scales only polynomially with distance. The proposed quantum repeater is composed from several basic known building blocks such as non-Gaussian operations for entanglement distillation and an iterative Gaussification protocol (for retaining the Gaussian character of the final state), but complemented with a heralded non-Gaussian entanglement swapping protocol, which allows us to avoid extensive iterations of quantum Gaussification. We characterize the performance of this scheme in terms of key rates for quantum key distribution and show a secure key can be generated over thousands of kilometers. 
\end{abstract}

\maketitle

\section{Introduction}

Today's society has ready access to more knowledge and information than at any time in our history. A key enabler of this is the internet which is underpinned by the worldwide telecommunications infrastructure. The way we currently process, manipulate and  transmit information is "classical" in nature, however with recent technological advances new paths are opening that allow us to exploit quantum mechanics and it principles \cite{Milburn1997,dowling2003,spiller2006}. Quantum communication \cite{kimble2008} and quantum computation \cite{feynman1982,Deutsch1985} are such examples where we can perform certain tasks that are either extremely hard or impossible with our classical hardware.  The most mature quantum information technology is known as quantum key distribution (QKD) and is a mechanism to establish secret communication between two remote parties \cite{gisin2002,lo2014}. Compared to traditional modern but classical cryptography implementations, it provides provable security based on the law of physics and not on the computational hardness of certain problems. QKD  requires a quantum communication channel between the two parties, but does not necessarily require challenging quantum operations necessary for large scale quantum computers. Consequently, devices for QKD have reached a high degree of experimental maturity with commercial products available \cite{IDQuantique} and long-time field tests have been conducted under real world conditions  ~\cite{jouguet2012field,sasaki2011field}. 

The majority of the QKD implementations are realized using weak coherent light / single photons with a discrete variable (DV) encoding such as polarization, path, time bin \cite{sangouard2011quantum}. However as the traditional telecommunication industry uses intense light fields, there is a strong possibility of incompatibility if one wants to use the existing network infrastructure for both classical and quantum applications. This can be overcome by using a continuous variable (CV) encoding into phase space degrees of freedom ~\cite{Weedbrook12,diamanti2015}. Since CV QKD only requires generation of Gaussian states and homodyne detection, it allows technologically simple, efficient and high frequency implementations. Regardless of whether DV or CV quantum states of light are used, QKD (and quantum communication in general) are severely limited in their communication distances by the exponential fibers losses \cite{sangouard2011quantum,munro2015}.

 Quantum repeaters (QR) are the natural solution to this issue ~\cite{Briegel1998,dur1999}, as they are considered the quantum analogue of signal amplifiers used in the conventional telecommunications industry. Various designs for single photon (DV based encoded) have been proposed in the last decade and their performance extensively  studied \cite{ duan2001,sangouard2011quantum,munro2012,muralidharan2015,munro2015}). The basic individual components for these repeaters have been implemented within a number of experimental efforts, yet their full integration has yet to be achieved.  

The continuous variable quantum repeater case is unfortunately the opposite. The field is still in its infancy as we do not even know whether a continuous variable quantum repeater (CV QR) is possible using polynomial resources. However given CV QKD's practical implementation advantages, it seems essential that we determine whether it's range limitation of approximately $100$ km or less (due to finite-size effects and excess noise) can be overcome using quantum repeaters ~\cite{jouguet2011}. Further complicating this is the fact that it is well known that a CV QR cannot solely be based on Gaussian operations \cite{namiki2014} due to the Gaussian entanglement distillation no-go theorem \cite{eisert2002,fiuravsek2002NoGo,giedke2002GaussianNoGo}. 
Non-Gaussian (NG) entanglement distillation protocols with subsequent Gaussification have been proposed \cite{browne2003,eisert2004,lund2009,fiuravsek2010,campbell2012,campbell2013} and shown to allow a larger degree of entanglement to be distributed than that expected from direct transmission \cite{campbell2013}. A preliminary CV quantum repeater scheme  \cite{dias2015} has been proposed (using noiseless linear amplifier~\cite{ralph2008NLA} based channel purification) that overcomes the exponential EPR fidelity scaling with distance but whose scaling with local operations is unknown.
 
In this article, we propose a CV quantum repeater scheme that distributes CV entangled states with arbitrary fidelity and a polynomial scaling in the distance. Our proposal builds on the NG entanglement distillation protocols \cite{lund2009,fiuravsek2010} and Gaussification protocols  \cite{browne2003,eisert2004} supplemented with a  heralded NG entanglement swapping operation, which allows to postpone the Gaussification protocol to the very end of the protocol. This avoids the need to run too many iterations of the Gaussificaiton protocol for which the convergence for practical situations is not known. There are a number of means by which we can analyze the performance of our QR scheme but the most natural here is in terms of CV QKD ranges and rates as this would be one of the first applications of such repeater. We need to show that it outperforms direct transmission for various distances. 

\section{Structure of the CV QR and its components}

\begin{figure}[b] 
\includegraphics[width=0.75\linewidth]{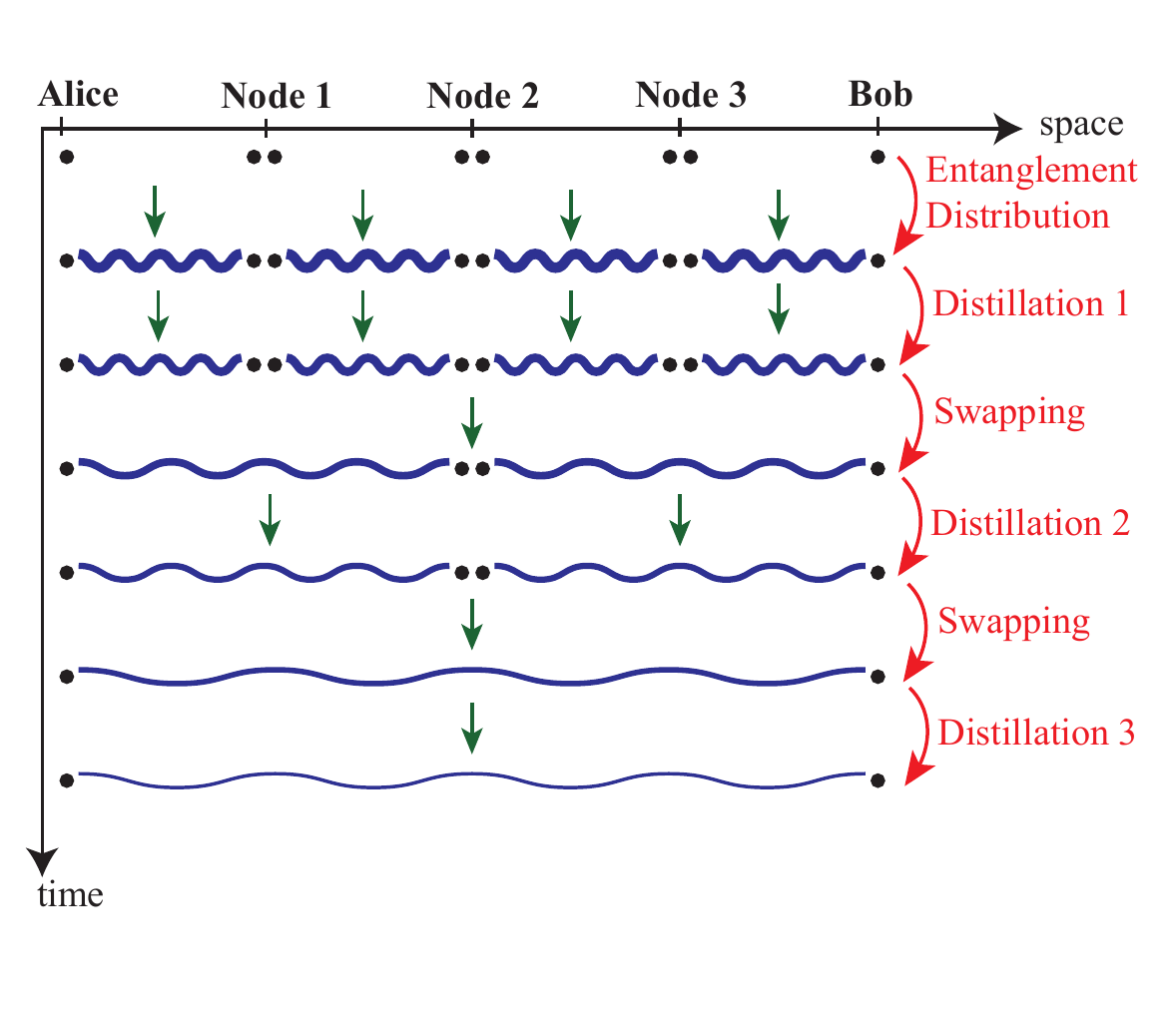}
\caption{Schematic illustration of a traditional first generation quantum repeater scheme \cite{munro2015} between Alice $A$ and Bob $B$ where that communication distance is divided into shorter distances by adding repeater nodes (stations) in between (three nodes in this case). The repeater scheme begins by establishing entanglement between the adjacent nodes. Then, entanglement distillation (Distillation 1) is performed to distill a smaller number of highly purified and entangled states (number of copies is illustrated by the thickness). Subsequently, entanglement swapping is performed until one established the required entangled state between Alice and Bob. Distillation 2 and Distillation 3 might be necessary to compensate the loss of purity and entanglement during swapping. Instead of using distillation protocols one can utilize a quantum error correction code for the distribution of the entanglement and for the errors induced by the swapping operation.  }
 \label{fig:QR}
\end{figure}

The basic structure of a CV quantum repeater (as illustrated in Fig.~\ref{fig:QR} and described in its caption) is similar to that used in the DV approach  \cite{Briegel1998,dur1999}. 
Each repeater station perform three types of operations: entanglement distribution (a technique to distribute entanglement between adjacent nodes), entanglement purification (a technique to increase the amount of entanglement shared between the nodes) and entanglement swapping (a technique to increase the range the entanglement is shared over). However the entanglement source  and the protocols for entanglement distillation (purification) and swapping vary crucially between the CV and DV cases. More specifically, our entanglement source is now a two mode squeezed state located between the adjacent repeater nodes (rather a source of Bell pairs of single photons), our purification protocols are non Gaussian entanglement distillation schemes (rather than simple qubit error detection codes) and our entanglement swapping schemes use homodyne based detection rather than probabilistic Bell state measurements. Further, the CV scheme uses a gaussification operation to return our distributed entangled state to approximately guassian after non-quassian operations have been performed on it (for example in entanglement distillation). In the next several sections we will describe these CV operations in detail. 

\section{Entanglement Source}

The most fundamental component  of any repeater scheme is the entanglement source and for the continuous variable scheme it is the two-mode squeezed vacuum (also known as the Einstein-Podolski-Rosen (EPR) state~\cite{epr35,Weedbrook12}) given by  
\begin{equation}\label{eq:EPR}
 \ket{ \mbox{\Large$\chi$}_{\lambda}} = \sqrt{1-\lambda^2} \sum_{k=0}^\infty \lambda^k \ket{k,k}. \, 
 \end{equation}
Here $\lambda=\tanh r \in [0,1)$ determines the strength of the squeezing (with r being the usual squeezing parameter). For $\lambda = 0$, we recover the vacuum, while for $\lambda\rightarrow 1$ we obtain the un-physical infinite energy state where both quadratures are perfectly correlated. Of course as we  need to distribute this entanglement between nodes channel losses are important. Here we will focus on the fully symmetric situation in which the losses for both modes of the entangled EPR states are symmetric meaning the source is placed in the middle between any adjacent repeater nodes.  

An EPR state with losses can conveniently be characterized in terms of its covariance matrix (CM). Because the state is Gaussian and the displacement is $0$, this determines the state uniquely. The CM of an  EPR state with symmetric losses from  transmission of each mode  through a  channel with transmissivity $\tau$ (losses 1-$\tau$) has the form
\begin{equation}\label{eq:CM}
\Gamma = \left( \begin{array}{cc}
C\idty & S \mathbb Z  \\
 S \mathbb Z  & C\idty 
 \end{array} \right) \,  ,
\end{equation}
where $\mathbb Z = \diag (1,-1)$, $C = 1 + \tau (\cosh(2 r) - 1) $ and $S=  \tau \sinh(2r)$. Next let us examine the protocols required for entanglement distillation and swapping. 

\section{Gaussification Protocol}\label{sec:Gaussification}

Our entanglement distribution process uses Gaussian states that retain this gaussian nature even under loss. However some of our entanglement distillation protocols result in non gaussian states being formed (see Appendix \ref{app:Gaussification}). This can be problematic and so a fundamental operation known as Gaussificaiton \cite{browne2003,eisert2004,campbell2012,campbell2013} is important. Gaussification protocols  allow us to make our states more gaussian in nature. 

\begin{figure} [htb] 
\includegraphics[width=0.8 \linewidth]{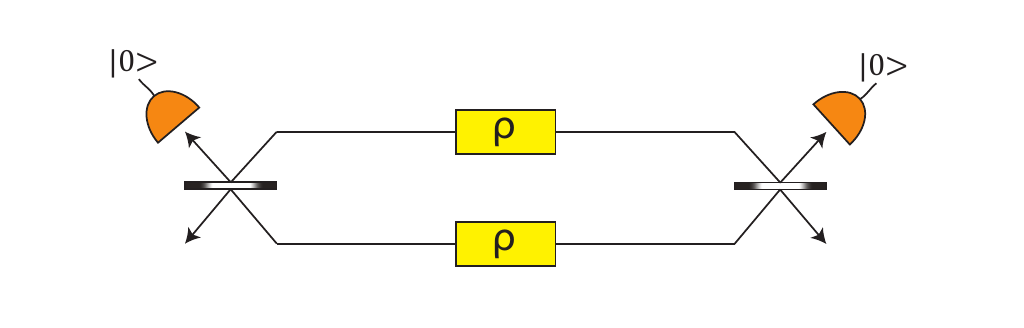}
\caption{Gaussification protocol: On both nodes, the two modes are mixed with a balanced BS. The operation succeeds conditioned on vacuum port detection (a Gaussian filtering operation) at both nodes. This procedure can be iteratively executed with the obtained state as the new input state until one is close enough to the desired Gaussian state. By choosing different filtering operations \cite{campbell2013} different Gaussification protocols are obtained (see Appendix \ref{app:GFO}).  }
 \label{fig:Gaussification}
\end{figure}

Consider that our initial resource state is given by the two-mode state $\rho^0$, where the two modes are at two space-like separated nodes $N$ and $M$. Then, in the first iteration of the Gaussification protocol  (see Fig.~\ref{fig:Gaussification}) two resource states $\rho^0\otimes\rho^0$ are used to generate a two mode state $\rho^1$ on $N$ and $M$ in a probabilistic but heralded fashion. In the second iteration, the same procedure is applied with the resource state $\rho^1$. Hence, from $\rho^1\otimes\rho^1$ a two-mode state $\rho^2$ is generated. In the limit of infinite iterations $i\rightarrow \infty$ one obtains a Gaussian state $\rho^\infty$, and an explicit formula how the CM can be calculated is known \cite{campbell2012,campbell2013}. Although, we make a strong distinction between NG entanglement distillation and Gaussification, we emphasize that generally the Gaussification protocol also increases the entanglement of the input state. 

\section{Entanglement distillation and purification protocols}

Entanglement distillation is the process by which we can take several copies of the quantum state and operate on them to obtain a new quantum state with increased entanglement. We are now going to  examine two distillation protocols that are both based on a probabilistic approach in which the desired states are distilled by conditioning on a suitable measurement outcome. The first one called symmetric photon replacement (PR)~\cite{browne2003,eisert2004} has the ability to efficiently increase the entanglement of the resulting state, but not its purity~\cite{fiuravsek2010}. The second protocol called purifying distillation overcomes this problem and allows to purify the state arbitrarily~\cite{fiuravsek2010}, however has the disadvantage that the success probability is very low and its implementation is much more demanding.

\subsection{Symmetric photon replacement distillation} 

\begin{figure}[htb]
\includegraphics[width=0.8 \linewidth]{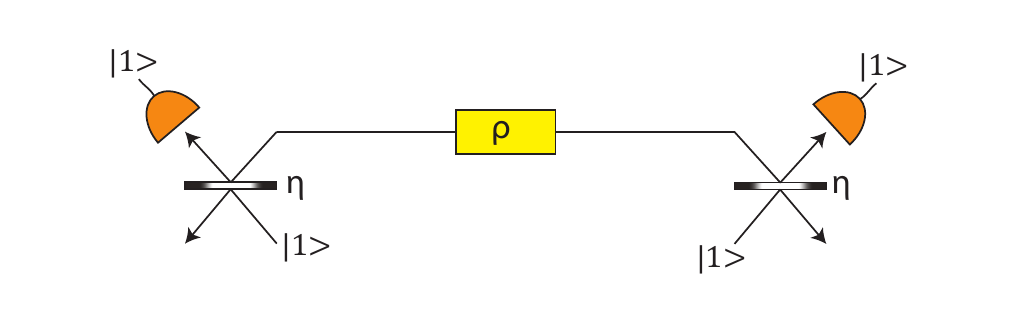}
\caption{In the symmetric PR distillation both modes are mixed with a single photon using a BS with transmissivity $\eta$. The operation is successful if a single photon is measured at the respective outcome of the BS. 
} \label{fig:PRdistillation}
\end{figure}

The starting point for this  distillation procedure shown in Fig.~\ref{fig:PRdistillation} is a two mode state $\rho$ shared between two nodes $N$ and $M$. In the symmetric PR distillation protocol, both modes are mixed at a beam splitter (appendix \ref{app:BS}) with transmittance $\eta$ with a single photon . The output port of the single photon is then measured with a single photon detector. The operation is successful if on both modes a single photon is detected and the corresponding output state is denoted by $\tilde\rho$.

At this stage we need to be able to characterize how effective our distillation has been. This is non trivial for a non Gaussian state but as detailed in Appendix (\ref{app:GFO},\ref{app:SPR}) for an initial state $\rho$, we can determine the  CM of the Gaussified sate $\rho^\infty$ quite easily if the low photon number matrix of $\rho$ in the Kronecker basis $\{\ket{0,0},\ket{0,1},\ket{1,0},\ket{1,1}\}$ has the form given in Equation~\eqref{eq:F1}. The Gaussified state is a lossy EPR state with squeezing parameter $\lambda=\lambda_\infty(\rho)$ and symmetric transmissivity $\tau= \tau_\infty(\rho) $ \cite{lund2009,fiuravsek2010}. We call $\tau_\infty(\rho)$ and $\lambda_\infty(\rho)$ the Gauss parameters of the state $\rho$. Since the low photon matrix of a symmetric EPR state has the form~\eqref{eq:F1} and it is conserved by the symmetric photon replacement (and all operations considered in here), the Gauss parameters provide a handy tool to characterize the effect of the distillation. It also enables us to define $\varepsilon(\rho) = \lambda_\infty(\rho) (1-\tau_\infty(\rho))$, which turns out to be equal to $\varepsilon(\rho)=\bra{1,0}\rho \ket{1,0} / \bra{1,1}\rho \ket{0,0}$~\cite{lund2009}. The symmetric photon replacement has now the simple property that $\epsilon(\tilde\rho) = \epsilon(\rho)$~\cite{lund2009}. 

However, as pointed out in~\cite{fiuravsek2010} any operation that leaves $\epsilon$ invariant cannot increase the purity of the state. In order to increase the purity $\epsilon$ has to be decreased. Hence, the disadvantage of symmetric PR distillation is that it will always decrease the purity of the resulting state~\cite{fiuravsek2010} as it increases entanglement. This can cause a problem for a CV QR as, for instance, subsequent application of entanglement swapping reduces the purity. For instance, with the symmetric PR distillation alone, we are not able to show that our CV QR scales for all distances only polynomial in the distance. 

\subsection{Purifying distillation.}\label{sec:PurifyingDist}

As the name suggests the purifying distillation protocol introduced in~\cite{fiuravsek2010} overcomes the problem of the symmetric PR distillation protocol and allows us to increase the purity of the state. 
\begin{figure}[htb]
\includegraphics[width=0.8\linewidth]{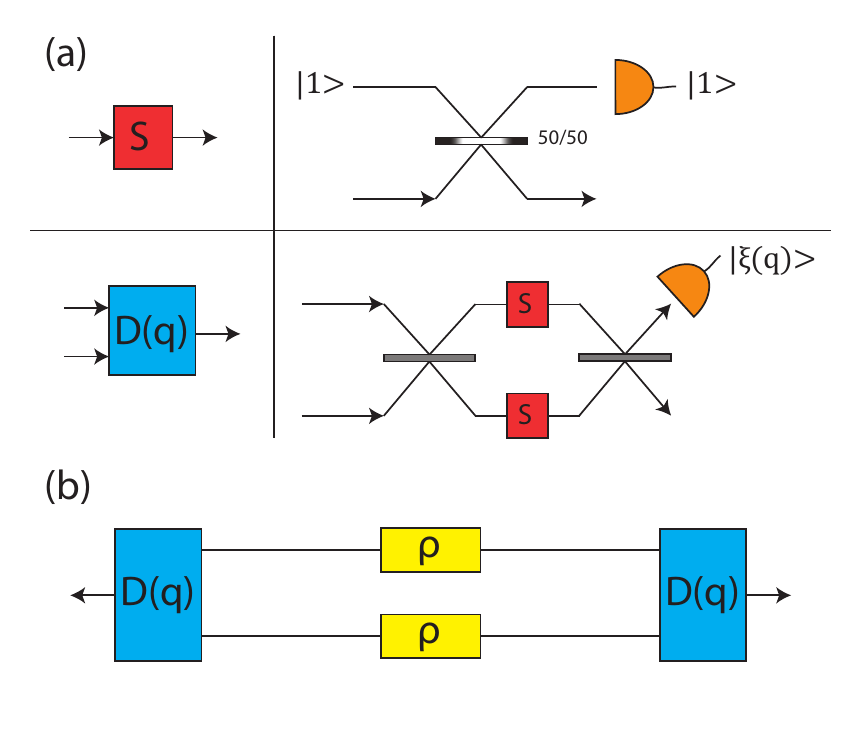}
\caption{The upper part (a) shows the schematic of the operation $S$ and $D(q)$ used in the purifying distillation protocol. $S$ is a photon replacement with a balanced BS. $D(q)$ is a probabilistic operation consisting of a Mach-Zehnder interferometer with $S$ placed in each paths and conditioning on $\ket{\xi(q)}$ on one of the output ports. The lower schematic (b) illustrates the purifying distillation operation in which $D(q)$ is applied to two initial states $\rho$. Heralding is applied on the successful application of $D(q)$ on both sides.  
}\label{fig:PurifyingDist}
\end{figure}
It however requires $2$ two-mode states $\rho$ as a resource in order to distil a two-mode state $\tilde\rho$ that has higher purity \footnote{Note that in~\cite{fiuravsek2010} it has been shown that the purity cannot be increased by simple operations on one two-mode state such as photon subtraction or by adding Gaussian post-selection.}. The structure of the protocol depicted in Fig.~\ref{fig:PurifyingDist} consists of a NG operations $D(q)$ performed on the two copies of the state $\rho$. The core operation $D(q)$, consists of a Mach-Zehnder interferometer in which a probabilistic NG operation $S $ is placed in both paths followed by a measurement projection onto the state
\begin{equation}
\ket{\xi(q)} =\frac{1}{\sqrt{1+q^2}}(q \ket{0} + \ket{1}) \,\,\,\,{\rm with}\,0\leq q \leq \infty.
\end{equation} 
Such a measurement can be implemented by displacements and photon addition together with projecting on the vacuum \cite{dakna1999}. Next critical to this $D(q)$ operation is the choice of $S$ and in this case we choose a  photon replacement operation with a balanced BS (see Fig.~\ref{fig:PurifyingDist}) due to its easy experimental implementation. This operation has the mathematical form 
\begin{equation}\label{eq:Sop}
S = \frac{1}{\sqrt {2^{\hat{n}+1}}} (\hat{n}-1) \, .
\end{equation} 
with $\hat{n}$ denoting  the number operator and filters out the single photon components. This approximates the operation $\hat n -\idty$ originally proposed by \cite{fiuravsek2010}  up to a factor depending on the photon number. 
The idea behind this choice is that it maps up to a phase $\bra{0}$ and $\bra{1}$ in the Heisenberg picture to $\bra{00}$ and $\bra{11}$, because the states $\bra{01}$ are filtered out by the Mach-Zehnder interferometer if in both arms the single photon events are filtered out. This has the consequence that $\tilde\rho_{ik,\alpha\beta} \sim \rho_{ik,\alpha\beta}^2$, which implies that $\varepsilon(\tilde\rho) = \varepsilon(\rho)^2$ (see Appendix \ref{app:SPR}). Hence, since $\varepsilon $ is smaller than $1$, $\varepsilon$ decreases which leads to Gauss parameters corresponding to a Gaussian state with increased purity (see Appendix  \ref{app:PD}).

\begin{figure}  [htb]
\includegraphics[width=0.8\linewidth]{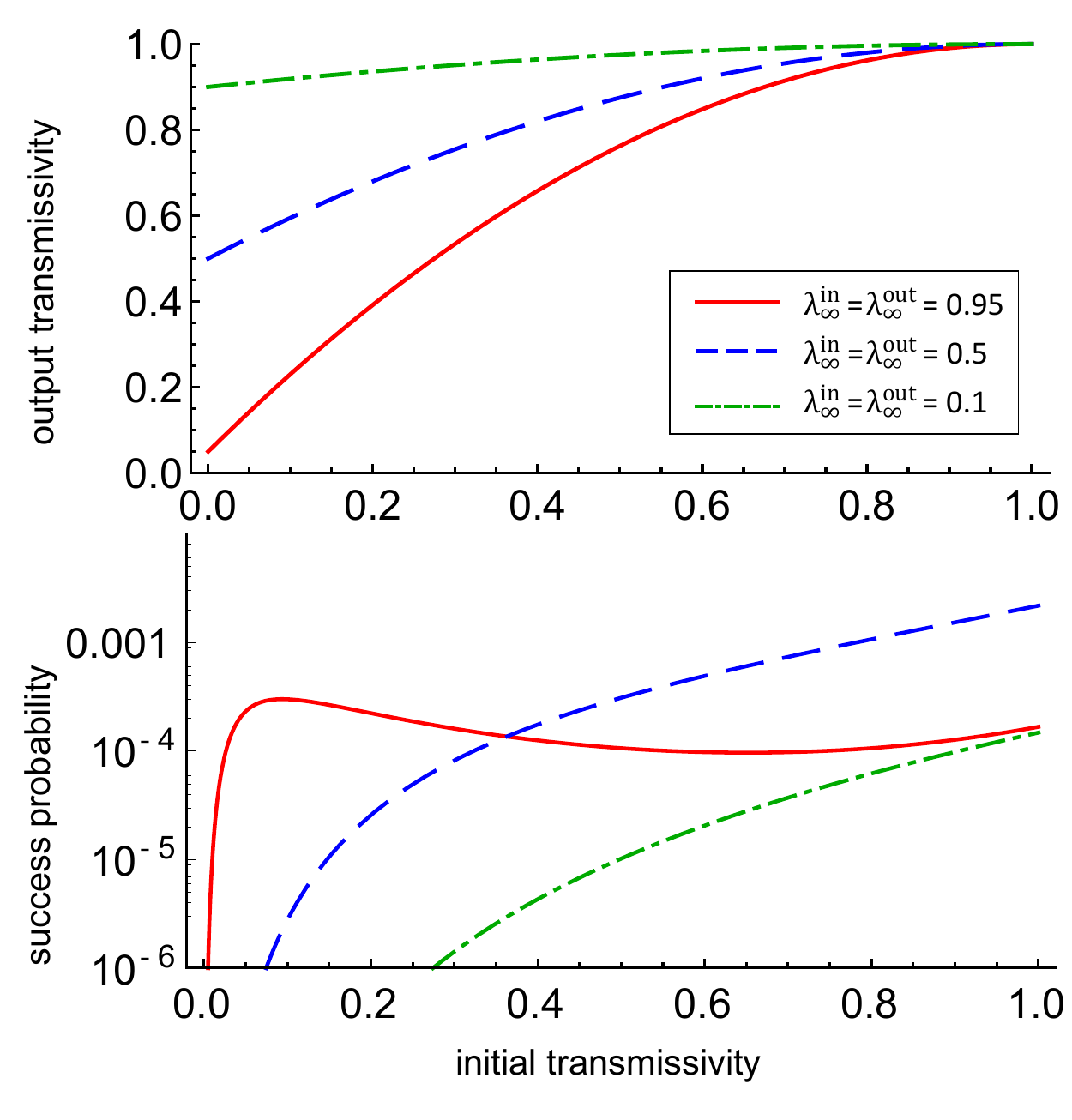}
\caption{The plot shows the Gauss parameter for the transmissivity of the output state $\tau_{\infty}(\tilde\rho)$ and the corresponding success probability $p_{\text{succ}}$ depending on the Gauss parameter for the transmissivity of the input state $\tau_\infty(\rho)$. Our projective measurement $\ket{\xi(q)}$ has the free parameter q depending on the choice of state we wish to condition on. We can tune $q$ such that the Gaussian parameter for the input squeezing $\lambda_\infty^{\text{in}}=\lambda_\infty(\rho)$ and output squeezing $\lambda_\infty^{\text{out}}=\lambda_\infty(\tilde\rho)$ are equal but  $\varepsilon(\tilde\rho)< \varepsilon(\rho)$, that is $\tilde\rho$ is more pure than $\rho$.}
\label{fig:PurifyingDistPlot}
\end{figure}

The implementation of $D(q)$ is more complicated than symmetric photon replacement distillation as it requires PR to be implemented twice within each repeater node as well as the measurement based projection onto the state (\ref{eq:Sop}). Further the success probability of the purifying protocol is much lower than PR. In particular, higher order photon terms are strongly suppressed. To illustrate this we plot in Fig.~\ref{fig:PurifyingDistPlot} the success probability of the purifying distillation as well as the Gauss parameter for the output transmissivity. 

\section{Entanglement swapping protocols} 

\begin{figure}  \includegraphics[width=\linewidth]{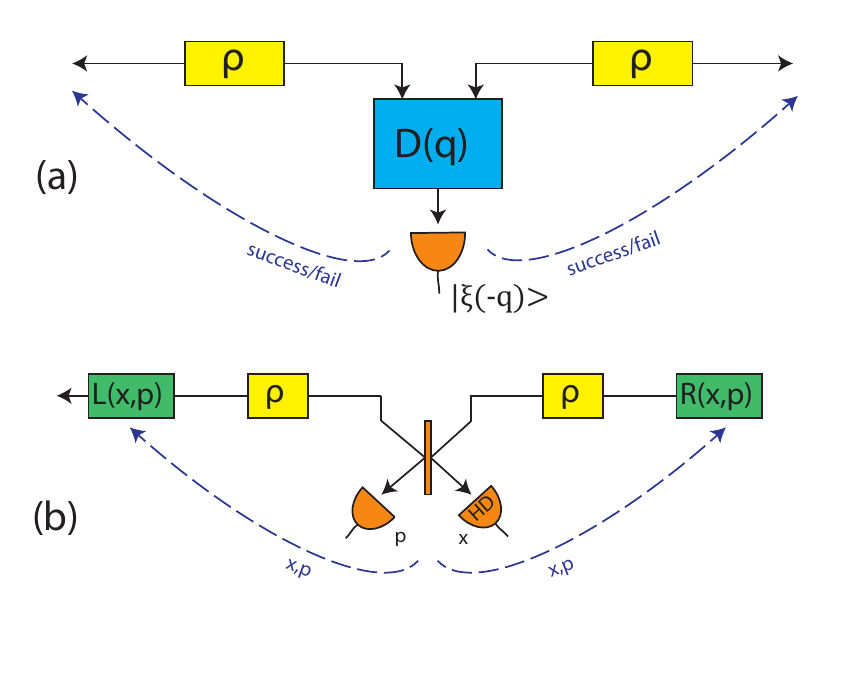}
\caption{ Schematic illustration (a) of the NG entanglement swapping protocol. The operation is heralded upon success of the operation $D(q)$ and measuring the state $\ket{\xi(-q)}$. (b) shows the schematics of the Gaussian entanglement swapping protocol. The two modes are mixed with a balanced BS and homodyne detection (HD) is used on both output ports to measure $X$ and $P$. A correction operation in form of a displacement depending on the measurement outcomes is applied on the left and right mode. 
} \label{fig:Swapping}
\end{figure}

We have now established entanglement distribution and distillation protocols for our CV quantum repater scheme. The final protocol required is entanglement swapping, which allows us to extend the range of our entanglement beyond that we can create between adjacent repeater nodes.  If the initial states are lossy EPR states one can simply chose the standard Gaussian swapping protocol ~\cite{hoelscher2011optimal} based on Gaussian teleportation~\cite{braunstein1998CVteleportation,ralph1998CVteleportation,vaidman1994CVteleportation}.  In such a case the two modes at the same node are mixed with a balanced BS, whereupon homodyne detection is used to measure the $X$ amplitude of one output and the $P$ amplitude of the other output (see Fig.~\ref{fig:Swapping} (b)). The measurement outcome are then send to the other nodes and a corresponding correction operation in form of a displacement is made. Given that the outcomes of the $X$ and $P$ measurement are $x$ and $p$, the displacements on the left mode is $g \sqrt{2} (-x +ip)$ and on the right mode $g\sqrt{2} (x +ip)$, where $g$ is the gain that has to be adjusted ~\cite{hoelscher2011optimal}. 

\subsection{Non gaussian entanglement swapping protocols} 

Before the Gaussian entanglement swapping can be applied, a Gaussification protocol has to be used to turn the NG state after entanglement distillation again into or close to a lossy EPR state. But since the above distillation protocols suppresses the higher photon number components, several iteration might be required to retain the states Gaussian character. This is especially a problem if one wants to distill a highly entangled state. An alternative NG entanglement swapping protocol is possible that does not require Gaussification. 

The idea is to swap the entanglement in the low photon number subspace spanned by the local one photon subspaces. For simplicity, let us consider the case with no losses, where the projection of an EPR state on to the local $1$-photon subspace is up to normalization given by $\ket{\tilde{\chi}_\lambda} = \ket{00} + \lambda \ket{11}$ (see Eq.~\ref{eq:EPR}). The tensor product is simply  
\begin{align}\label{eq:approxEPRtensor}
\ket{\tilde{\chi}}_{12}\otimes\ket{\tilde{\chi} }_{34} &  = \ket{00}_{14} \ket{00}_{23}   + \lambda^2 \ket{11}_{14} \ket{11}_{23}  \\ 
& \quad +  \lambda ( \ket{01}_{14} \ket{01}_{23} + \ket{10}_{14}  \ket{10}_{23}) \, , \nonumber
\end{align}
where modes $2$ and $3$ are assumed to be at the same node. Hence, in order to swap the entanglement we have to project modes $2$ and $3$ onto a state proportional to $\ket{\tilde\chi_a}$ to obtain 
\begin{equation}
 \ket{00} + a \ket{11} \, .
\end{equation}
Now letting $a=1/\lambda$ we  perfectly swapped our initial truncated EPR states. In order to realize a projection onto a state $\ket{\tilde\chi_a}$ with experimentally feasible operations, we need to cut out the components $\ket{01},\ket{10}$. However, as we have seen in Section~\ref{sec:PurifyingDist} this can be achieved by a Mach-Zehnder interferometer with photon replacement in the two paths, that is, an operation similar to $D(q)$. Indeed, a straightforward computation shows that 
\begin{align}\label{eq:NGswap1}
D(q)^*\ket{\xi_{\bar q}} = \frac{q\bar q \ket{00} -1/4 \ket{11}}{\sqrt{2(1+q^2)(1+\bar{q}^2)}}  \, .
\end{align}
Hence, by choosing $\bar q = -q$ we obtain the projection on a state proportional to $\ket{\tilde\chi_a}$. This motivates our choice of the NG swapping operation consisting of an application of $D(q)$ followed by condition on the state $\ket{\xi_{-q}}$ as illustrated in Fig.~\ref{fig:Swapping} (a). Note that the protocol is not deterministic and the information whether or not it succeeded has to be communicated to the other nodes. This is in contrast to the Gaussian entanglement swapping, which is deterministic. 

\begin{figure} 
 \includegraphics[width=0.8\linewidth]{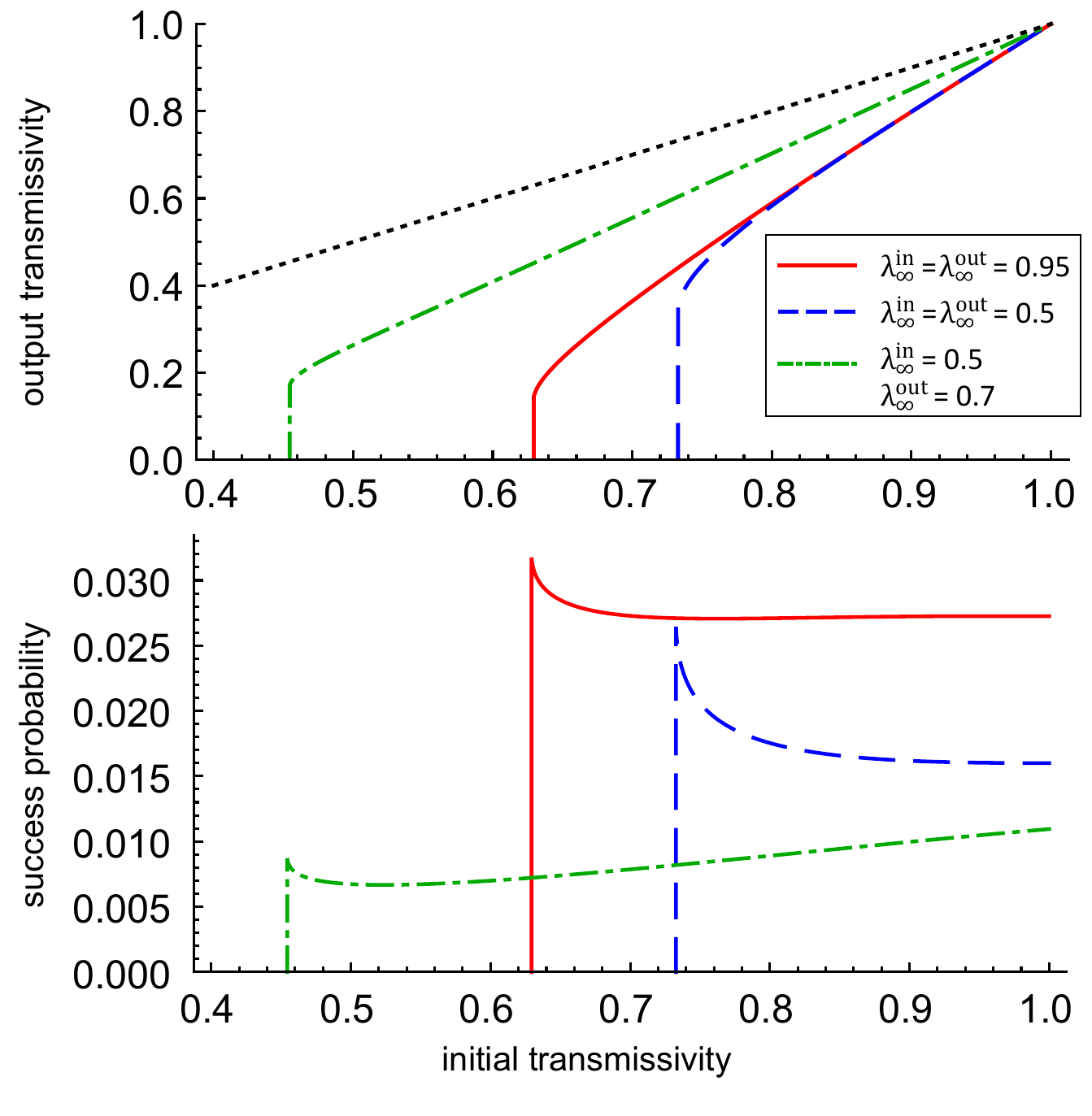}
\caption{Pot of the  Gauss transmissivity parameter for the output state $\tau_{\infty}(\tilde\rho)$ and the corresponding success probability $p_{\text{succ}}$ depending on the  input state Gauss transmissivity parameter $\tau_\infty(\rho)$. The Gaussian parameters for the input squeezing $\lambda_\infty^{\text{in}}=\lambda_\infty(\rho)$ and output squeezing $\lambda_\infty^{\text{out}}=\lambda_\infty(\tilde\rho)$ are shown. 
In order to illustrate the decrease of the Gauss  output transmissivity parameter (upper plot), we also plotted the input transmissivity (dotted line). We see that there exists a minimal initial transmissivity for the protocol to conserve the desired structure of the input state (i.e., the Gaussification protocol converges). The success probability increases and the threshold for the initial transmissivity gets reduced for larger squeezing $\lambda_\infty^{\text{in}}=\lambda_\infty^{\text{out}}$. The threshold for the initial transmissivity is also reduced if the output squeezing $\lambda_\infty^{\text{out}}$ is larger than initial squeezing $\lambda_\infty^{\text{in}}$, but the success probability decreases.}\label{fig:NGswapPlot}
\end{figure}

We can now characterize the effect of the NG swapping protocol on a state $\rho$.  By using~\eqref{eq:NGswap1} 
the resulting state $\tilde \rho$ in matrix form  (appendix \ref{app:ES}) is given by
\begin{align}
\tilde\rho_{ij,\alpha\beta} & = \frac{1}{2(1+q^2)^2} \, \Big( q^4 \ \rho_{i0,\alpha 0}\ \rho_{0j,0\beta}+ \frac 14\  \rho_{i1,\alpha 1} \  \rho_{1j,1,\beta} \nonumber \\
 & \quad  + \frac{q^2}{2} \ (  \rho_{i0,\alpha 1} \ \rho_{0j,1\beta} + \rho_{i1,\alpha 0} \ \rho_{1j,0\beta})  \Big) \, .
\end{align} 
where the free $q$ parameter can be adjusted to give the desired Gauss squeezing parameter $\lambda_\infty$.
In Figure \ref{fig:NGswapPlot} we plot the success probability and Gauss output transmissivity parameter for for different values of the Gauss parameters for the input squeezing and the output squeezing.  We observe that there exists a threshold for the input Gauss transmissivity parameter $\tau_\infty(\rho)$ below which the operational interpretation of the Gauss parameters fails, that is, the Gaussification of the output state does not converge any more. According to Fig.~\ref{fig:NGswapPlot},this threshold crucially depends on the Gauss parameter for the initial and final squeezing. More precisely, if the Gauss parameter for the output squeezing is larger then the input squeezing, a lower Gauss parameter for the transmissivity is possible. But at the same time this decreases the success probability of the NG swapping protocol as well as the purity of the output state.

\section{Quantum Repeaters for CV QKD} 

We now have all the components (entanglement distribution, entanglement distillation and entanglement swapping) required to design and quantitatively analyze how a CV quantum repeater would work. We do however need to specify a figure of merit for its performance. This will obviously depend on the application, but  as a natural first step we will estimate the secret key-rate if the distributed state had be used for CV QKD using a homodyne based detection scheme ~\cite{diamanti2015}. In order to keep the calculation of the key rate simple we use the asymptotic key rate formula that is uniquely determined by the CM of the distributed state \cite{lodewyck2007,Weedbrook12,diamanti2015}. It is important to note that the state does not have to be a Gaussian state in order for the key rate formula to apply due to the extremality property of Gaussian states~\cite{Cerf2006GaussOpt,Navascues2006GaussOpt}. Moreover, in~\cite{leverrier2015} it has been shown how the CM can be estimated without assuming that the state is Gaussian. 

To begin let us examine how the key rate can be computed without CV QR restricting ourselves for simplicity to only collective attacks. If the input state is independent and identically given by $\rho$, the key rate in the asymptotic limit is \cite{lodewyck2007,circac09}
\begin{equation} \label{eq:KeyRate}
  r(\rho) =  I(X_A:X_B)_{\rho^{G}} - I(X_B:E)_{\rho^G}  \, . 
\end{equation}
where $\rho^G$ denotes the Gaussian state with the same second moments as $\rho$, $I(X_A:X_B)_{\rho^G}$ the mutual information of Alice's and Bob's key generating measurement applied to the Gaussian state, $I(X_B:E)_{\rho^G}$ the mutual information (or Holevo quantity) between the eavesdropper and Bob's measurement. 

We emphasize that the key rate above is for the case of reverse reconciliation, which is favourable for high losses. $r(\rho)$ can then be applied to calculate the key rate for direct transmission. For that, we choose $\rho$ as an EPR state where one mode is sent through a fibre channel with transmittance $\tau(l) = 10^{-l \mu/10}$ where $l$ is the channel length and $\mu$ the loss per kilometer (set to $0.2/$km in the case). For an asymmetric loss distribution using reverse reconciliation \cite{grosshans2003}, the key rate for CV QKD decreases linearly with the transmittance, and thus, exponentially with the distance. This expression also gives us a direct way to compare our repeaters performance. 

For our CV quantum repeater protocol the distributed entangled states are symmetric in nature rather than the optimal asymmetric loss distribution used in the original CV QKD. This has the consequence that a positive key rate cannot be obtained for low losses and an EPR state with relatively high transmissivity is required. In order to compute this key rate, let us assume that we distribute the state $\sigma$ using a CV QR. Then, the key rate associated with the CV QR is then 
\begin{equation}
r_{\QR}(\sigma) = \frac 1{N_{QR}(\sigma)} r(\sigma)  \, , 
\end{equation} 
where $N_{\QR}(\sigma)$ denotes the average number of initial EPR states that are consumed to distribute the state $\sigma$ with the CV QR.  The normalized key rate $r(\sigma)$ is the important quantity to analyze the practicality under finite-size effects and excess noise due to the homodyne detection. However if $r(\sigma)$ is sufficiently large, this means that even under finite size effects and additional noise a positive key rate can be obtained. In fact, for direct transmission the key rate $r(\rho)$ gets so low for distances above about $100$km that under practical conditions a positive key rate is no longer possible \cite{diamanti2015}. 

\subsection{Scalings: Overcoming exponential loss}\label{sec:ExpScal}

The key step for our CV quantum repeater scheme is to show that in principle it can overcome the exponential decay of the key rate in the distance for direct transmission, rather than worrying about whether the performance is optimal or not. Given this, let us investigate a CV QR based on the NG entanglement swapping protocol, rather than the Gaussian entanglement swapping as it avoids several further Gaussification steps. Consider that the total distance we want to establish our key over is $L$ and it is divided into $2^n$ equal segments each of length $l=L/2^n$ with a repeater station joining each segment. In total we have $N=2^n-1$ repeater stations (excluding the end points) and $n$ swapping operations are required. Since we are examininig a symmetric QR protocol, the source is placed in the middle between any two adjacent repeater stations and so the initial transmittance of each mode is $\tau_0=\tau(l/2)$.  We further fix the final squeezing $\lambda_{\text{final}}$ and transmittance $\tau_{\text{final}}$ of the target EPR state that we want to distribute between the end nodes. 

Our scheme, as depicted in Figure (\ref{fig:QR}), begins with the distribution of the EPR state between adjacent nodes. Once this has been successful, we apply entanglement distillation protocols (symmetric PR and purifying distillation) to generate a state that has Gauss parameters that are at least those of the desired target state. The success probability $p_0$ of the initial distillation depends on the initial squeezing and the distance between adjacent nodes $l$, and is thus, independent of the total distance.  Subsequently, NG entanglement swapping is applied that succeeds with probability $p_{\text{swap}}$, which depends on the Gauss parameter of the input state (i.e. determined by the target state). Another round of entanglement distillation is used to retain again a state that complies with the target state. The success probability $p_{\text{dist}}$ for the entanglement distillation only depends on the properties of the NG swapping protocol and the squeezing and transmittance of the target state.

In the same way NG swapping and entanglement distillation can simply be applied as many times as required to reach the desired distance $L$. Hence, we find that the success probability for the swapping is polynomial in the number of swapping operations $n$, that is, $(p_{\text{swap}} p_{\text{dist}})^n$~\footnote{Note that due to our swapping and distillation protocol, we only set a minimal condition for the Gauss parameters determined by the target state. Hence, the actual probability may slightly vary but can be bounded by optimizing over all states that comply with the target state}.  Finally, the Gaussification protocol is applied to generate an approximate Gaussian state with the squeezing and transmittance. Given that the success probability for the Gaussification protocol is $p_{\text{Gauss}}$, we find that the overall success probability  
\begin{align}
p_{\text{tot}}&  = p_0 p_{\text{Gauss}} (p_{\text{swap}} p_{\text{dist}})^n  
=  p_0 p_{\text{Gauss}} \left(\frac{L}{l}\right)^{\log_2(p_{\text{swap}} p_{\text{dist}})} \, 
\end{align} 
scales polynomial in the total distance $L$. 
Applied to CV QKD, the above CV QR leads to a key rate of $r_{\QR} = p_{\text{tot}}\ r(\rho_{\text{target}})$, where $\rho_{\text{target}}$ denotes the target state.

\subsection{Performance}

\begin{figure} [htb]
\includegraphics[width=0.8\linewidth]{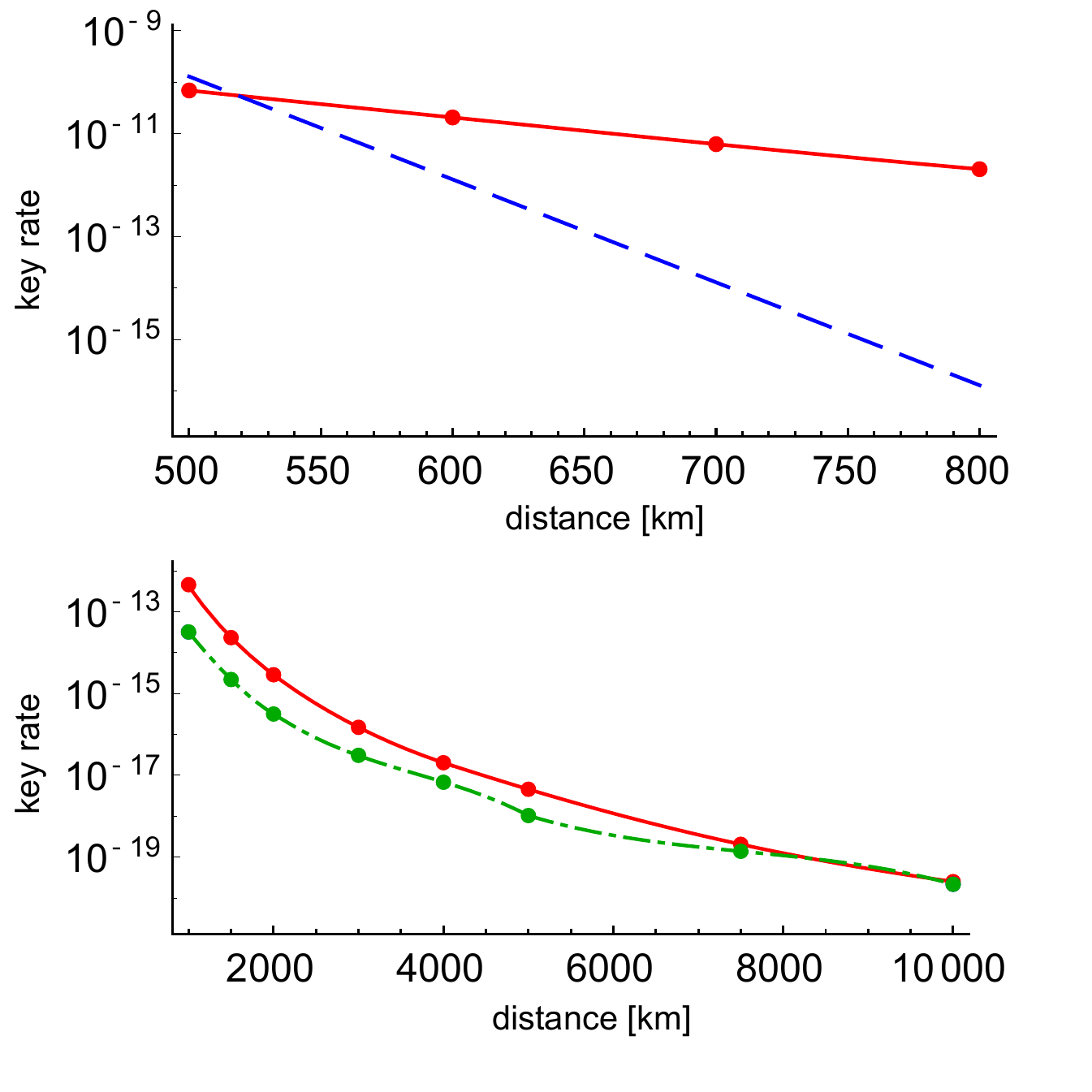}
\caption{Plot showing the performance of the CV QR based on NG entanglement swapping. The upper plot compares the optimal key rate of the CV QR ithout additional purifying distillation (straight line), with the key rate for direct transmission (dashed). We see that the crossing point is slightly above 500 km and the gap increases fast with the distance. The lower plot compares the key rate of the CV QR without (straight) and with (dashed) additional purifying entanglement distillation. We observe that while for low distances the key rate without additional purifying distance is significantly larger, the gap closes for large distances above $10000$ km. The dots correspond to actual simulation points and the curves are interpolations through it. }
\label{fig:QRNGswap}
\end{figure}

While the scaling of our CV repeater scheme could be established in a simple fashion, determining its performance requires us to make assumptions about our physical devices and their imperfections. For simplicity, we will assume an ideal system here where our only source of error are losses in the channel. We will assume perfect single photon detectors with a detection efficiency of $1$ and zero dark counts. With these assumptions  the key rates are plotted and compared with direct transmission in Fig.~\ref{fig:QRNGswap}.  The dots correspond to the numerically optimized key rates and the curve is an interpolation through it. We see that the crossing point where the CV QR scheme performs better than direct transmission is near $500$ km.  

Now the optimal CV QR strategy for the distances considered in Fig.~(\ref{fig:QRNGswap}) is different to the one used in Section~\ref{sec:ExpScal} to show the polynomial scaling in the distance. This is due to the fact that the success probability for the purifying distillation, and thus, $p_{\text{dist}}$ is very low (see Fig~\ref{fig:PurifyingDistPlot}). Instead, the optimal strategy for all practical distances considered here is to first distribute a state with very low squeezing ($\lambda \leq 0.05$) and apply a symmetric PR distillation to distil a state with Gauss parameter $\lambda_\infty \approx 1$. The reason is that for low squeezing the decrease in purity due to fiber losses can be minimized (a higher initial squeezing would result in a very mixed state after the distribution). Since our distributed state has high purity, it is not necessary to apply a purifying distillation protocol and a symmetric PR distillation is sufficient. Indeed, if the symmetric PR distillation is applied to boost the Gauss parameter to $\lambda_\infty\approx 1$, also the transmittance gets boost to  $\tau_{\infty} \approx 1$. The price we have to pay is simply a low success probability of our symmetric PR distillation, which is still favorable compared to applying additional purification. The fact that a low initial squeezing is beneficial in combination with symmetric PR distillation has already been noted in the context of Gaussian entanglement distillation~\cite{lund2009}.

A second strategy is similar to the previous one except that we add one additional purifying distillation after the first swapping operation. We see in Fig.~\ref{fig:QRNGswap} that the obtained key rate is slightly lower than with no additional purification for the plotted distances. However, the gap becomes closer as larger the distance gets and finally after about $10000$km a crossing is expected. Moreover, we expect that the first strategy without additional purifying distillation may not allow distribution over arbitrary distances as the NG entanglement swapping will fail as soon as the Gauss parameter $\tau_\infty$ falls below a certain threshold (see Fig.~\ref{fig:NGswapPlot}). 

\begin{table}[b]
\begin{center}
\begin{tabular}{|l||r|r|r||r|r|r|}
\hline
 & $\lambda_\infty$ & $\tau_\infty$ &  $p_{\text{succ}}$ & $\lambda_\infty$ & $\tau_\infty$ &  $p_{\text{succ}}$ \\ \hline   
Initial State &  $0.013$  & $0.013$ &  & $0.04$ & $0.11$ &  \\ \hline  
PR Dist. &  $0.95$  & $0.99$ & $6.5\cdot 10^{-8}$ & $0.98$ & $0.96$ &  $4.5\cdot 10^{-5}$\\ \hline   
1st Swap &  $0.95$  & $0.97$ & $0.027$ & $0.98$ & $0.93$ & $0.028$  \\ \hline   
Purification &    &  &  & $0.98$ & $0.995$ & $2.5\cdot 10^{-3}$  \\ \hline   
2nd Swap & $0.9$  & $0.94$    & $0.028$ & $0.98$ & $0.99$ & $0.028$ \\ \hline   
3rd Swap &   $0.65$ & $0.83$ & $0.035$  & $0.98$ & $0.98$ & $0.038$ \\ \hline   
4th Swap &    &  &  &  $0.55$ & $0.91$ & $0.038$ \\ \hline   
Gaussification &  $0.65$  &  $0.83$& $0.093$ &  $0.55$ & $0.91$ & $0.11$  \\ \hline   
\end{tabular}
\end{center}
\caption{Table showing the Gauss parameters $\lambda_\infty$ and $\tau_\infty$ at the different stages of the QR protocol as well as the success probability $p_{\text{succ}}$ of the operations. The total distance is $1500$ km and the values of the first three columns correspond to the strategy without purification, while the right three columns show the one with applying the purifying distillation. The initial state refers to the EPR state after distirbution. We see that the first strategy needs only $3$ swapping operations ($7$ repeater stations), while the optimal strategy with purification requires $4$ swapping operations ($15$ repeater stations). We note that only a slightly lower key rate is obtained for the protocol with purification and only $3$ swapping operations such that it might be favourable from a practical point of view. In the last swapping operation it is beneficial to decrease $\lambda_\infty$, which is favourable to a high normalized key rate $r(\sigma)$ if only $2$ iterations of the Gaussification protocol are applied. The key rate $r_\QR(\sigma)$ and the normalized key rate $r(\sigma)$ without (with) purification are $2.3\times 10^{-14}$ ($2.2\times 10^{-15}$) and $0.14$ ($0.27$).} 
\label{table:Parameters}
\end{table}

In our simulation, we optimized over the initial, final and intermediate Gauss parameter for the squeezing and the number of swapping operations. In Table~\ref{table:Parameters}, we give example values of the optimal Gauss parameters for $1500$ km and both strategies. We see that the final state distributed over the total distance has much higher purity for the second strategy, but the success probability is lower. In both cases we see that normalized key rates is relatively high, which implies that it is robust against finite-size effects and excess noise. Moreover, since the distributed states provide a high normalized key rate, the scheme is robust against finite-size effects. So, even for distances below $500$ km our scheme proves to be useful for practical CV QKD. 

\section{Discussion and Outlook}\label{sec:Conclusion}

In this work we have presented a CV QR scheme that has a success probability scaling polynomial in the distance and can distribute entangled EPR states with arbitrary squeezing and fidelity over arbitrarily long distances. These distributed EPR states can then be used for many CV communication applications such as  quantum teleportation or CV QKD. Alternatively they can also  be used for communicating DV quantum information used in a qubit based quantum computer~\cite{takeda2013}. We analyzed the performance of our protocol by examining its use in CV QKD (see Fig.~\ref{fig:QRNGswap}) and found the crossing point where the key rates with the CV QR is larger than with direct transmission is near $500$km. By analysing the key rate for larger distances, we observe the strong positive effect of the polynomial scaling. This is due to the fact that the purification protocol has a low success probability. Indeed the lower plot in Fig~\ref{fig:QRNGswap} shows that the range where an application of the purifying distillation protocol is beneficial is beyond $10'000$km. This illustrates one of the current problems of the CV QR, which is the low efficiency of the purifying entanglement distillation protocol. Compared to the entanglement increasing symmetric PR distillation, it has a success probability of about two to three magnitudes lower. Hence, an important step in improving the proposed CV QR protocol is to find entanglement distillation protocols that increase the purity, but have a practical success probability. It is known that purification is not possible with simple heralded operations on only one mode~\cite{fiuravsek2010}. But we still expect that under moderate increase of the experimental difficulty an improved purification scheme is possible. 

This is not the only challenge that remains for CV QR, in fact their are many of them. One of the most important is given in the situation where the final state has to be very close to a Gaussian EPR state. Since the Gaussification protocol is based on a non-commutative central limit theorem, we expect that the approximation error $\epsilon$ goes as ${1}/{\sqrt{N}}$, where $N$ is the number of iterations. Hence, given the iterative scheme of the Gaussification protocol the success probability scales exponential in $1/\epsilon^2$. This problem could been greatly avoided in our application to CV QKD, where a Gaussian state is not essential.  But it is not clear how a NG or (badly) approximate Gaussian state performs in a Gaussian teleportation or Gaussian entanglement swapping protocol. This is also the reason why we only obtained a quantitative estimation if using the NG entanglement swapping protocol, and could only approximate the scaling of the CV QR with the Gaussian entanglement swapping protocol (see Fig.~\ref{fig:QRCVswap}). However, since Gaussian entanglement swapping has the big advantage to be deterministic, it is important to determine the approximation errors after a finite number of iterations of the Gaussification protocol and how such errors effect the output-fidelity of the state after Gaussian entanglement swapping. 

Even if all these issues are overcome, the performance of these CV QR's will be quite slow due to the long range classical communication that is needed in the entanglement distillation steps. What we have designed here is analogous to the first generation DV QR's which were also extremely slow \cite{sangouard2011,munro2015}. Improvement in DV repeater design to second \cite{jiang2009,munro2010}  and third \cite{fowler2010,munro2012,muralidharan2014} generations offered the potential for huge speed improvements. The second generation DV schemes replaced the traditional purification algorithms with full deterministic local node error corrections \cite{jiang2009,munro2010} and so their operational speed was limited by the communication times between adjacent nodes rather than the communication between the end nodes. Orders of magnitude increase in performance have been predicted \cite{jiang2009,munro2010}. The third generation DV QR went further and replaced the heralded entanglement distribution with a deterministic (or near deterministic) approach using photon loss coding \cite{ralph2005,kok2007}. The performance of such schemes were thus only limited by the gate times within the repeater nodes, and so GHz rates have been predicted \cite{fowler2010,munro2012,muralidharan2014}. If we can move the second and third generation DV QR concepts across to the CV QR, then we would expect large gains in our performance. This however requires the development of new CV error correction codes (both for gates errors and channel loss). 

The main application we have discussed in this CV QR work is CV QKD and we have found in principle that the CV QR are an advantages in terms of communication rate once our users are more than 500 km apart. We must emphasize however that our CV QR protocol should also be useful for CV QKD for lower distances. In practical implementations the distance of CV QKD is limited to less than $100$ km due to finite-size effects and excess noise~\cite{jouguet2011}. With our CV QR scheme the generated key rates are practical and robust to finite-size effects and additional excess noise induced by homodyne detection. Further an extension to an asymmetric entanglement distribution scheme for our CV QR protocol may further improve the distance.

{\bf Now to conclude}, our CV QR scheme allows us to distribute entangled EPR states with an arbitrary fidelity over long distances that has a success probability scaling polynomial with distance. Further we have shown how the distributed states can be used for CV QKD with rates that exceed the ones by direct transmission. Our approach here has been to show that CV QR are in principle possible and  so we have assumed perfect single photon sources and detectors and not included the effected of realistic errors / noise (apart from channel loss). While we expect that our CV QR scheme is robust under small deviation of these assumptions, it is important to make this more precise in the future work. In particular, the question whether single photon sources can be replaced by weak coherent states is crucial to allow for high frequency CV QR implementations. This would provide a clear advantage over DV QR approaches which often rely on true single photon sources. 

\emph{Acknowledgements.---} We gratefully acknowledge useful discussions with Jens Eisert, Akihito Soeda, Koji Azuma and Kiyoshi Tamaki.

\bibliography{libraryCVRepeater}

\begin{appendix}

\section{Non Gaussian state representation}\label{app:Gaussification}

As  we have mentioned earlier, a CV QR has to include NG operations and so the resulting NG states can no longer be fully described by the CM. Instead we need a different parametrization. It is useful to use the matrix coefficients of the state in the photon number basis. For any two-mode state $\rho $, we denote the matrix elements in the number basis by
\begin{equation}
\rho_{k l, \alpha \beta} = \bra{k,l} \rho \ket{\alpha,\beta} \, 
\end{equation}
where $\{\ket{n}\}$ denotes the number basis of a single mode. A  lossy EPR state satisfies the constraints
\begin{eqnarray} \label{eq:MatrixElemEPRCond}
\rho_{k l, \alpha \beta} &=& \rho_{l k,  \beta \alpha} \, \\ 
\label{eq:MatrixElemEPRCond1}
\rho_{k l, \alpha \beta} &=& 0 \, , \, \,  \text{if} \, \,   k-\alpha \neq l-\beta \, , \\
\label{eq:MatrixElemEPRCond2}
\rho_{k l, \alpha \beta} &\in&  \mathbb{R} \, , 
\end{eqnarray}
and the explicit form of the coefficients for low photon numbers can be found in, e.g.,~\cite{lund2009}.

\section{Gaussification operation} \label{app:GFO}

In section \ref{sec:Gaussification} (and depicted Figure \ref{fig:Gaussification}) we discussed Gaussification based on vacuum port detection at each node \cite{browne2003,eisert2004}. This  Gaussian filtering operation was considered because it allowed a simple characterization of the CM of the emerging Gaussian state by the low photon matrix elements of $\rho^1$, that is, the resulting state after the first iteration. The characterization of the CM is even simpler if the low photon matrix in the Kronecker basis $\{\ket{0,0},\ket{0,1},\ket{1,0},\ket{1,1}\}$
\begin{equation}
F_1(\rho^0):= (\rho^0_{kl,\alpha\beta})_{k,l,\alpha,\beta=0}^1 \, ,
\end{equation}
has the form 
\begin{equation}\label{eq:F1}
F_1(\rho) = \left( \begin{array}{cccc}
\rho_{00,00} & 0 & 0 & \rho_{11,00}  \\
 0  & \rho_{01,01} & 0 & 0 \\ 
 0 & 0 & \rho_{01,01} & 0 \\ 
 \rho_{11,00} & 0 & 0 & \rho_{11,11} 
 \end{array} \right) \,  . 
\end{equation}
Note that this is the case for the symmetric EPR state (\ref{eq:MatrixElemEPRCond}-\ref{eq:MatrixElemEPRCond2}). Then, the CM is determined by the low photon number matrix $F_1(\rho)$ only and 
the CM of the corresponding Gaussified state has the same form as an EPR state with symmetric losses~\eqref{eq:CM}.
 In order to express the CM in terms of $\rho_{kl,\alpha \beta}$ it is convenient to introduce the quantities 
\begin{align}
 \varepsilon(\rho) & = \frac{\rho_{10,10}}{\rho_{11,00}}  \;\;\;\;\;\;\;\; \Lambda(\rho) = \frac{ \rho_{11,00}}{\rho_{00,00}} \, . 
\end{align}
We then find that the CM of $\rho^\infty$ is given by~\eqref{eq:CM} with~\cite{lund2009} 
\begin{align}
C & = \frac{ \Lambda^2(1-\varepsilon^2) + 1 }{(1-\varepsilon \Lambda)^2 - \Lambda^2} \;\;\;\;\;S  = \frac{2\Lambda }{(1-\varepsilon\Lambda)^2- \Lambda^2} \, ,
\end{align}
where $\varepsilon=\varepsilon(\rho)$ and $\Lambda=\Lambda(\rho)$. We need to point out explicitly that there exists states $\rho$ for which the obtained CM is not physical, that is, it does not satisfy the necessary condition $\Gamma + i\Omega\geq 0$ with $\Omega$ the symplectic form \cite{Weedbrook12}. In this case, the Gaussification protocol does not converge. 

We can now invert the expressions $C = 1 + \tau (\cosh(2 r) - 1) $ and $S=  \tau \sinh(2r)$ from Equation \eqref{eq:CM} to express the state $\rho^\infty$ as a lossy EPR state with squeezing parameter $\lambda=\lambda_\infty(\rho)$ and symmetric transmissivity $\tau= \tau_\infty(\rho) $ given by 
\begin{align}\label{eq:lambdainf}
\lambda_\infty(\rho)  = \varepsilon(\rho) + \Lambda(\rho) (1-\varepsilon(\rho)^2) \\ 
 \tau_\infty(\rho)  = (1-\varepsilon(\rho)^2) \Lambda(\rho)/\lambda(\rho) \, . \label{eq:tauinf}
\end{align} 
which yields~\cite{lund2009}
\begin{equation}
\varepsilon(\rho) = \lambda_\infty(\rho) (1-\tau_\infty(\rho)) \, .
\end{equation}

\section{Symmetric PR distillation} \label{app:SPR}
We now have the tools to  analyze the performance of a NG operation in terms of $\lambda_\infty$ and $\tau_\infty$ for the hypothetical Gaussified state.  Let us assume that the input state $\rho$ satisfies~\eqref{eq:F1} and let us denote the output state obtained by symmetric PR distillation by $\tilde\rho$.   Using the expansion of the BS operation given in Appendix~\ref{app:BS}, we find that 
\begin{equation}\label{eq:CoeffPRdist}
\tilde\rho_{kl, \beta \gamma} = \bar\alpha_{k,1|0}\bar\alpha_{l,1|0}\alpha_{\beta,1|0}\alpha_{\gamma,1|0} \rho_{kl,\beta\gamma} \, ,
\end{equation} 
where $\alpha_{ij|t}$ is defined in~\eqref{eq:BScoeff}. While $\bar c$ denotes the complex conjugate of $c$, we emphasize that in our case the BS coefficients $\alpha_{ij|t}$ are real. 
The formula above shows immediately that conditions (\ref{eq:MatrixElemEPRCond}-\ref{eq:MatrixElemEPRCond2}) are preserved by the symmetric PR distillation, and thus, also the structure of $F_1(\rho)$ is conserved.  

Hence given our constraints are satisfied, we can characterize the symmetric PR distillation by analysing the Gauss parameters of the input and output state, i.e., the relation between $(\lambda_\infty(\rho),\tau_{\infty}(\rho))$ and $(\lambda_\infty(\tilde\rho),\tau_{\infty}(\tilde\rho))$. As shown in~\cite{lund2009}, the symmetric PR distillation can be characterized by 
\begin{align}\label{eq:PRdistEps}
\varepsilon(\tilde\rho) &= \varepsilon(\rho) \\ 
\Lambda(\tilde\rho) & = \beta(\eta)^2 \Lambda(\rho)\, ,\label{eq:PRdistLambda}
\end{align}
where $\Lambda(\rho)= \rho_{11,00}/\rho_{00,00}$ and $\beta(\eta)= (2\eta^2-1)/\eta$. By tuning $\eta$ the Gauss parameter $\tilde\lambda_\infty$ can be made arbitrary close to $1$ but at the expense of decreasing the success probability. Care also needs to be taken to ensure that $\eta$ is not chosen such that $\Lambda_\infty(\tilde\rho) > 1/(1+\epsilon)$, otherwise $\tilde\lambda_\infty>1$ which is unphysical. This means that the Gaussification protocol does no longer converge.

\section{Purifying distillation details} \label{app:PD}

Due to the importance of the purifying distillation, let us examine this scheme in more details by characterizing the effect of the operation on the matrix elements in the number basis. These are computed by
\begin{align}
\tilde\rho_{kl,\alpha\beta} =  \bra{k,l} D(q)\otimes D(q) (\rho\otimes \rho) D(q)^* \otimes D(q)^* \ket{\alpha,\beta} \, .
\end{align}
A straightforward calculation using~\eqref{eq:Sop} shows
\begin{equation}
D(q)^* \ket{k} = \sum_{l=-1}^k \ket{k-l}\otimes (q d_{k|l} \ket{l} + \tilde d_{k|l}\ket{l+1}) \, 
\end{equation} 
where 
\begin{align}
d_{k|l} & = \gamma_k(q) \sum_{t=0}^k (k-t-1)(t-1)\alpha_{k0|t}\alpha_{(k-1)t|(l-t)} \\ 
\tilde d_{k|l} & = \gamma_k(q) \sum_{t=-1}^k \sqrt{\frac 12} t(k-t-1) \alpha_{k1|t} \alpha_{(k-t)(t+1)|(l-t)} \, 
\end{align}
with $\gamma_k(q) = (1/2)^{(k+1)/2}/\sqrt{1+q^2}$ and where we use the convention $d_{k|-1}=0$. Let us now assume that $F_1(\rho)$ has the form in~\eqref{eq:F1}. This allows us explicitly compute $\tilde\rho_{kl,\alpha\beta}$ and determine the coefficients of $F_1(\tilde\rho)$ as
\begin{align}
\rho_{00,00} & = \frac{1}{2^2} \frac{q^4}{(1+q^2)^2} \,   \rho_{00,00}^2 \\
\rho_{10,10} & =  \frac{1}{2^4} \frac{q^2}{(1+q^2)^2} \,  \rho_{10,10}^2 \\
\rho_{01,01} & =  \frac{1}{2^4} \frac{q^2}{(1+q^2)^2} \,  \rho_{01,01}^2 \\
\rho_{11,00} & =  \frac{1}{2^4} \frac{q^2}{(1+q^2)^2} \,  \rho_{11,00}^2 \\
\rho_{11,11} & =  \frac{1}{2^{10}} \frac{1}{(1+q^2)^2}\,  \rho_{11,11}^2  \, ,
\end{align}
with all the others are $0$. Hence, the structure of $F_1$ is conserved by the purifying distillation. 

Numerical calculations further suggests that the condition (C2) is conserved in general by the purifying distillation. Unfortunately, we were not able to prove this explicitly. However for our CV QR it is sufficient that the structure of $F_1$~\eqref{eq:F1} is conserved, in order for the Gauss parameter to remain meaningful. This is because for all sub-protocols $F_1(\tilde\rho)$ is only a function of $F_1(\rho)$. Using the above formulas, we see that for the purifying distillation~\cite{fiuravsek2010}
\begin{align}
\varepsilon(\tilde\rho) = \varepsilon(\rho)^2 \, . 
\end{align}
Since for meaningful states $\varepsilon\leq 1$, we find that $\varepsilon$ reduces with purifying distillation leading to state with increased purity.

\section{NG entanglement swapping}\label{app:ES}

In the main text we showed that the matrix elements of the resulting state $\tilde \rho$ were given by 
\begin{align}
\tilde\rho_{ij,\alpha\beta} & = \frac{1}{2(1+q^2)^2} \, \Big( q^4 \ \rho_{i0,\alpha 0}\ \rho_{0j,0\beta}+ \frac 14\  \rho_{i1,\alpha 1} \  \rho_{1j,1,\beta} \nonumber \\
 & \quad  + \frac{q^2}{2} \ (  \rho_{i0,\alpha 1} \ \rho_{0j,1\beta} + \rho_{i1,\alpha 0} \ \rho_{1j,0\beta})  \, . 
\end{align} 
Under the condition that $F_1(\rho)$ has the form~\eqref{eq:F1}, we then obtain that the non-zero terms of $F_1(\tilde\rho)$ are given by 
\begin{align*}
\tilde\rho_{00,00} & = \frac{1}{2(1+q^2)^2} (q^4 \rho_{00,00}^2 + \frac 14 \rho_{10,10}^2 ) \\ 
\tilde\rho_{01,01} & = \frac{(1)}{2(1+q^2)^2} (q^4\rho_{01,01}\rho_{00,00} + \frac 14 \rho_{01,01}\rho_{11,11} ) \\ 
\tilde\rho_{10,10} & = \frac{(1)}{2(1+q^2)^2} (q^4\rho_{10,10}\rho_{00,00} + \frac 14 \rho_{10,10}\rho_{11,11} ) \\
\tilde\rho_{11,00} & = \frac{(1)}{4(1+q^2)^2} q^2\rho_{11,00}^2 \\ 
\tilde\rho_{11,11} & = \frac{(1)}{2(1+q^2)^2} (q^4\rho_{10,10}\rho_{01,01} + \frac 14 \rho_{11,11}^2 ) \, .
\end{align*}
Hence, also the NG swapping protocol NG swapping protocol conserves the essential structure of our state, that is,  the low photon number matrix $F_1(\rho)$ remains its form~\eqref{eq:F1}. 
The free parameter $q$ is conveniently adjusted in the way that it fixes the Gauss parameter for the squeezing $\lambda_\infty$.

\section{Beamsplitter in the Fock basis} \label{app:BS}

The BS transformation with transmittance $\tau$ transforms the ladder operators $a,b$ as $a \mapsto \sqrt{\tau} a + \sqrt{1-\tau} b $ and $  \mapsto \sqrt{\tau} a - \sqrt{1-\tau} b $. Since the BS conserves the photon number its action on a photon number eigenstate can be written as 
\begin{equation}\label{eq:BS}
U^{\BS} \ket{k,l} = \sum_{t=-l}^{k} \alpha_{k,l|t} \ket{k-t,l+t} \, . 
\end{equation} 
A straightforward calculation then leads to 
\begin{align} \label{eq:BScoeff} 
\alpha_{k,l|t} & = \sqrt{\frac{(k-t)!(l+t)!}{k! l! }} \sum_{m = \max\{0,-t\}}^{\min\{l,k-t\}} (-1)^{l-m}  \\
 & \quad \quad  \times \binom{k}{m+t} \binom{l}{m} \sqrt{\tau^{k+l-2m-t} (1-\tau)^{2m +t}} \, . \nonumber
\end{align}  

\end{appendix}

\end{document}